\documentclass[lettersize,journal]{IEEEtran}
\usepackage{amsfonts,dsfont,amssymb,amsbsy,amsmath,paralist,theorem,bm,ifthen,color}
\usepackage[pdfstartview=FitH,bookmarksnumbered,unicode,bookmarksopen=true]{hyperref}
\usepackage{graphicx}
\usepackage{epstopdf}
\usepackage{multirow}
\usepackage{subfigure,relsize}
\usepackage{mathrsfs}
\usepackage{bm}
\usepackage{stfloats}
\usepackage{url}
\usepackage[noadjust]{cite}
\usepackage{cases,booktabs}
\usepackage{graphicx,algorithmic,algorithm,relsize}
\usepackage[justification=centering]{caption} 
\usepackage[table]{xcolor}

\newtheorem{lem}{Lemma}
\newtheorem{prop}{Proposition}

\newtheorem{coro}{Corollary}

\newcommand\pin{\ensuremath{{\rm Pin}}}

\newcommand\hb{\ensuremath{{\bf h}}}

\newcommand\Xb{\ensuremath{{\bf X}}}
\newcommand\xb{\ensuremath{{\bf x}}}

\newcommand\Vb{\ensuremath{{\bf V}}}
\newcommand\vb{\ensuremath{{\bf v}}}

\newcommand\psib{\ensuremath{{\bm \psi}}}

\newcommand\wo{\ensuremath{{\textrm{wo}}}}
\newcommand\w{\ensuremath{{\textrm{w}}}}

\newcommand\E{\ensuremath{{\mathbb{E}}}}

\newcommand\SNR{\ensuremath{{\rm SNR}}}

\newcommand\Nset  {\ensuremath{{\mathcal{N}}}}
\newcommand\Mset  {\ensuremath{{\mathcal{M}}}}

\newcommand\st    {\ensuremath{{\rm s.t.}}}

\graphicspath{{fig/}}

\definecolor{green}{RGB}{34	195	46}
\definecolor{red}{RGB}{220 0 0}

\usepackage{graphicx} 

\title{Pinching-Antenna System Design with LoS Blockage: Does In-Waveguide Attenuation Matter?}

\author{ Yanqing Xu, \IEEEmembership{Member, IEEE,}
         Zhiguo Ding, \IEEEmembership{Fellow, IEEE,}
         Octavia A. Dobre, \IEEEmembership{Fellow, IEEE,}\\
         and Tsung-Hui Chang, \IEEEmembership{Fellow, IEEE}
         \thanks{\smaller[1] Y. Xu is with the School of Science and Engineering, The Chinese University of Hong Kong, Shenzhen, 518172, China (email: xuyanqing@cuhk.edu.cn).}
         \thanks{\smaller[1] Z. Ding is with the University of Manchester, Manchester, M1 9BB, UK (email: zhiguo.ding@manchester.ac.uk).}
         \thanks{\smaller[1] O. A. Dobre is with the Faculty of Engineering and Applied Science, Memorial University, St. Johns, NL A1C 5S7, Canada (e-mail:odobre@mun.ca).}
         \thanks{\smaller[1] T.-H. Chang is with the School of Artificial Intelligence, The Chinese University of Hong Kong, Shenzhen, 518172, China (email: changtsunghui@cuhk.edu.cn).} 
         \thanks{\smaller[1] This work has been submitted to the IEEE for possible publication. Copyright may be transferred without notice, after which this version may no longer be accessible.}
        \vspace{-0mm}}

\date{\today}

\begin{document}

\maketitle

\begin{abstract}
    In the literature of pinching-antenna systems, in-waveguide attenuation is often neglected to simplify system design and enable more tractable analysis. However, its effect on overall system performance has received limited attention in the existing literature.
    While a recent study has shown that, in line-of-sight (LoS)-dominated environments, the data rate loss incurred by omitting in-waveguide attenuation is negligible when the communication area is not excessively large, its effect under more general conditions remains unclear. This work extends the analysis to more realistic scenarios involving arbitrary levels of LoS blockage. We begin by examining a single-user case and derive an explicit expression for the average data rate loss caused by neglecting in-waveguide attenuation. The results demonstrate that, even for large service areas, the rate loss remains negligible under typical LoS blockage conditions.
    We then consider a more general multi-user scenario, where multiple pinching antennas, each deployed on a separate waveguide, jointly serve multiple users. The objective is to maximize the average sum rate by jointly optimize antenna positions and transmit beamformers to maximize the average sum rate under probabilistic LoS blockage. To solve the resulting stochastic and nonconvex optimization problem, we propose a dynamic sample average approximation (SAA) algorithm. At each iteration, this method replaces the expected objective with an empirical average computed from dynamically regenerated random channel realizations, ensuring that the optimization accurately reflects the current antenna configuration. Extensive simulation results are provided to the proposed algorithm and demonstrate the substantial performance gains of pinching-antenna systems, particularly in environments with significant LoS blockage.

\end{abstract}


\section{Introduction} 
Multi-antenna technologies have been a cornerstone in the evolution of wireless communication systems, supporting critical capabilities, such as massive connectivity, high spectral efficiency, and improved link reliability \cite{xu2025distributed,larsson2014massive}. By leveraging spatial diversity and advanced beamforming techniques, multi-antenna architectures, particularly massive multiple-input multiple-output (MIMO), have dramatically improved the performance of cellular and local area networks \cite{bjornson2023twenty}. Despite these advances, conventional multi-antenna systems are typically built upon fixed-location antenna arrays with static spatial configurations, which inherently limits in their ability to adapt to rapidly changing environments. In particular, in the presence of user mobility and dynamic obstacles that introduce random line-of-sight (LoS) blockages, these fixed-location antenna deployments often struggle to maintain robust and high-throughput links, leading to degraded system performance and reliability.

To address the inherent disadvantages of traditional fixed-position antennas, recent years have witnessed significant advances in flexible and reconfigurable antenna technologies, such as reconfigurable intelligent surfaces (RISs) \cite{tang2020wireless}, movable antennas \cite{zhu2023movable}, and fluid antennas \cite{wong2021fluid}. These innovative paradigms aim to enhance wireless systems with greater adaptability by dynamically modifying the spatial configuration or physical position of antenna elements in response to environmental changes. By leveraging such spatial reconfigurability, these solutions offer the potential to mitigate the effects of user mobility, environmental dynamics, and other adverse propagation conditions. Nevertheless, a key limitation persists: these technologies are generally unable to modify the large-scale characteristics of the wireless channel or to efficiently establish new LoS links. For example, while RIS can manipulate the phases of reflected signals, it offers limited ability to overcome severe LoS blockage or shadowing. Similarly, the adjustment range of movable and fluid antennas often have a limited operational range, typically spanning only a few wavelengths, which constrains their ability to reposition effectively in the presence of dynamic LoS blockages. As a result, these techniques may be insufficient to dynamically reconstruct LoS links or maintain robust performance across extended communication regions, particularly in complex or highly dynamic environments.

Pinching antennas, which utilize radiating elements coupled to dielectric waveguides, have recently gained increasing attentions as a promising candidate for flexible antenna designs \cite{suzuki2022pinching}. This architecture enables the antenna element to traverse a broad spatial region along the waveguide, facilitating real-time adjustment of its location within the communication area \cite{ding2024flexible}. This dynamic reconfigurability allows the system to proactively reposition the antenna toward favorable locations, e.g., next to the served users, thereby establishing and maintaining strong LoS communication links, even in the presence of significant obstacles or rapidly changing environmental conditions. This spatial flexibility greatly enhances the system’s ability to mitigate blockages, accommodate user mobility, and sustain high-throughput, reliable communications. As a result, the pinching-antenna system demonstrates inherent adaptability and robustness, making it as a compelling candidate for enabling robust wireless access in next-generation communication networks.

Although pinching-antenna systems offer promising advantages, their study is still in its formative phase. In particular, how key parameters, including antenna placement, waveguide properties, and blockage characteristics, affect overall system performance remains insufficiently explored. To address this gap, this work presents a comprehensive analysis of pinching-antenna system performance under practical constraints, with a focus on understanding the influence of in-waveguide attenuation and random LoS blockage.

\subsection{Related Works}
The research community has increasingly turned its attention toward the unique properties of pinching-antenna systems, investigating their application potential and associated design challenges across a wide range of scenarios \cite{ding2024flexible,xu2025rate,xie2025low,wang2025modeling,bereyhi2025downlink,zhou2025gradient,tyrovolas2025performance,ding2025los,wang2025pinching,wang2025antenna,xu2025qos,ding2025pinching,ouyang2025rate,mao2025multi,ding2025analytical}. In particular, \cite{ding2024flexible} conducted a comprehensive study on the use of pinching antennas within various system architecture, including signal and multiple waveguide systems. The study highlighted how spatial reconfigurability of pinching antennas can substantially improve the ergodic capacity compared to traditional fixed antennas. From the perspective of data rate maximization, \cite{xu2025rate} formulated a pinching-antenna position optimization problem and proposed a low-complexity two-stage algorithm that simultaneously reduces large-scale path loss and ensures constructive signal superposition at the user. In multiuser and multi-waveguide scenarios, pinching antennas have been integrated within MIMO architectures \cite{wang2025modeling,bereyhi2025downlink}, where joint optimization of beamforming and antenna positioning has led to substantial gains in spectral efficiency. The outage performance of pinching-antenna system under random user locations was analyzed in \cite{tyrovolas2025performance}, providing insights into system reliability in stochastic environments. For more practical environments where LoS paths may be obstructed, the performance of pinching-antenna systems under blockage was investigated in \cite{ding2025los,wang2025pinching}. 
Meanwhile, the constraint of supporting a single stream per waveguide has motivated exploration into non-orthogonal multiple access (NOMA)-based solutions for pinching-antenna systems \cite{wang2025antenna,xu2025qos}. Recent efforts have also extended pinching-antenna research to emerging paradigms such as integrate sending and communications (ISAC) \cite{ding2025pinching,ouyang2025rate,mao2025multi}. Specifically, the work \cite{ding2025pinching} utilized a Cram\'{e}r-Rao lower bound (CRLB)-based metric to quantify the positioning accuracy achieved by pinching antennas. The capacity region of a pinching-antenna-enabled ISAC system was characterized in \cite{ouyang2025rate}, while \cite{mao2025multi} studied joint optimization of antenna positions and beamformers in multi-waveguide ISAC system designs.

While significant progresses have been made in the investigation of pinching-antenna systems, a key aspect of their physical-layer modeling is often simplified. Specifically, the end-to-end channel of a pinching-antenna system inherently includes an in-waveguide attenuation component, arising from signal propagation through the dielectric waveguide \cite{xu2025pinching}. 
This simplification is often made to enable tractable analysis and efficient algorithm design. However, the implications of this modeling choice have received limited attention. One exception is \cite{xu2025pinching}, which investigated the impact of in-waveguide attenuation under a LoS-dominated environment and found that the performance degradation is insignificant when the service area is not excessively large.
Nevertheless, practical wireless environments are more complex. In particular, the LoS links may not always available due to the obstacles and user movement. Therefore, it is necessary to investigate the system performance by explicitly considering both in-waveguide attenuation and probabilistic LoS blockage. 

\subsection{Contributions}
To obtain more practical understandings of in-waveguide attenuation in realistic wireless environments, this work considers a general system model that explicitly incorporates this probabilistic LoS blockage phenomenon into the system modeling of pinching-antenna systems. Comparing to the literature, such an extended framework provides a comprehensive and insightful perspective on system behavior in non-ideal propagation conditions. The contributions of this paper are as follows:
\begin{itemize}
    \item We begin by studying a single-user pinching-antenna system under probabilistic LoS blockage, with a focus on quantifying the impact of in-waveguide attenuation on achievable average data rate. We derive closed-form expression for the data rate degradation caused by neglecting this attenuation when optimizing antenna placement. This expression explicitly characterizes the dependence of the rate degradation on the key system parameters, including the attenuation coefficient, waveguide height, blockage density, and communication area size. The analysis reveals two key insights. First, in sparse obstacle blockage environments, the rate loss grows with the square of the waveguide length, suggesting that segmenting the waveguide into shorter sections can mitigate the effect of in-waveguide attenuation. Second, in dense obstacle blockage scenarios, the rate degradation converges to a constant that is typically negligible even in a large communication area, indicating that attenuation can be reasonably ignored in such settings. These findings provide practical guidelines for waveguide deployment and highlight the conditions under which the simplified model remain valid.
    
    \item We then extend our investigation to a multi-user MIMO downlink system, where multiple single-antenna users are served via multiple pinching antennas, each activated on an individual waveguide. The objective is to maximize the average sum rate by jointly designing beamforming vectors and antenna placements under random LoS blockage. This joint optimization is particularly challenging due to the coupling between variables and the stochastic nature of LoS blockage. 
    To tackle this challenge, we develop a dynamic sample average approximation (SAA) framework that approximates the expectation over blockage patterns using empirical averages from regenerated channel realizations at each iteration.  Within this framework, an alternating optimization approach is adopted, where beamformers are updated via the weighted MMSE method \cite{shi2011iteratively}, and antenna positions are refined using a particle swarm optimization (PSO) algorithm \cite{kennedy1995particle}.
\end{itemize}

Extensive simulations are provided to validate the performance of pinching-antenna systems in the presence of random LoS blockage and proposed algorithms under various settings. The results demonstrate that the dynamic SAA-based algorithm efficiently solves the challenging joint optimization problem and consistently outperforms a zero-forcing (ZF) beamforming-based baseline in terms of system sum rate. Moreover, the pinching-antenna systems achieve significant performance gains over fixed-antenna systems in the presence of LoS blockage. This highlights the system's ability to dynamically reposition antennas to maintain favorable propagation conditions.

The paper is organized as follows. Section \ref{sec: single user} introduces the single-user system model and characterizes the average rate loss from ignoring in-waveguide attenuation. Section \ref{sec: multiple user} presents the multi-user system and the dynamic SAA-based optimization strategy. Section \ref{sec: simulation} reports the simulation results, and Section \ref{sec: conclusion} concludes the paper.


{\bf Notations:} Throughout this paper, lowercase and uppercase bold letters (e.g., $\xb$ and $\Xb$) are used to denote column vectors and matrices, respectively. 
The set of $N$-dimensional complex vectors is denoted by $\mathbb{C}^{N}$. The superscript $(\cdot)^\top$ refers to the transpose, and $||\xb||^2$ represents the squared Euclidean norm.
The notation $\mathcal{R}\{C\}$ indicates the real part of a complex number $C$, and $\E_n[\cdot]$ denotes the expectation over the random variable $n$.
A Bernoulli random variable $x \sim \mathrm{Bernoulli}(p)$ has success probability $p$, while $x \sim \mathcal{U}(a,b)$ denotes a continuous uniform distribution on the interval$[a,b]$. The notation $\int_a^b f(x)  dx$ indicates the definite integral of $f(x)$ over $[a, b]$.

\section{Pinching-Antenna System Design in Single-User Case} \label{sec: single user}

This section focuses a communication setup, where a single pinching antenna is deployed along a waveguide to serve a user equipped with a single antenna. Without loss of generality, we assume the user’s position is randomly located within a square region of side length $D$. the user's $3$D coordinates be $\psib = [\bar x, \bar y, 0]$, where $0 \leq \bar x \leq D$ and $-\frac{D}{2} \leq \bar y \leq \frac{D}{2}$. The waveguide is placed above the user region, aligned along the $x$-axis at height $d_v$. Its feed point is located at $\bm{\psi}_0 = [0, 0, d_v]$, and the position of the pinching antenna on the waveguide is denoted by $\widetilde\psib^{\pin} = [\tilde x, 0, d_v]$.

\subsection{Channel Model with LoS Blockage and In-waveguide Attenuation}
Taking the in-waveguide attenuation into consideration, the LoS channel between the pinching antenna and the user is given by \cite{xu2025pinching}
\begin{align}\label{eqn: channel model siso}
    &h^{\text{LoS}} = \frac{\eta^{\frac{1}{2}}e^{-j \left( \frac{2\pi}{\lambda} \|\bm{\psi} - \boldsymbol{\widetilde{\psi}}^{\pin}\| + \frac{2\pi}{\lambda_g} \|\bm{\psi}_0 - \boldsymbol{\widetilde{\psi}}^{\pin}\| \right)}}{\|\bm{\psi} -\boldsymbol{\widetilde{\psi}}^{\pin}\|e^{\alpha \|\bm{\psi}_0 - \boldsymbol{\widetilde{\psi}}^{\pin}\|}},
\end{align}
where $\eta = \frac{c^2}{16\pi^2 f_c^2}$ is a constant based on the speed of light $c$ and carrier frequency $f_c$, $\alpha$ is the in-waveguide attenuation coefficient, and $\lambda, \lambda_g$ are the free-space and waveguide wavelengths, respectively. 
The value of $\alpha$ varies depending on waveguide structure and frequency and is typically around $0.08$ dB/m \cite{bauters2011planar}. 

In practical wireless environments, the presence of obstacles, such as buildings and furniture, as well as the user movement introduces randomness in the environment, causing the existence of a LoS link between the antenna and the user to become probabilistic.
To capture this effect, we consider a realistic scenario that incorporates LoS blockage. 
Specifically, the presence or absence of a LoS path is modeled as a Bernoulli random variable $\gamma \in \{0,1\}$, where $\gamma = 1$ represents the presence of a direct LoS path and $\gamma = 0$ indicates its absence. 
The probability that an LoS link exists is modeled as \cite{bai2014analysis}
\begin{align} \label{eqn: los probability}
    \mathbb{P}(\gamma = 1) = e^{-\beta \|\boldsymbol{\widetilde{\psi}}^{\pin}-\boldsymbol{\psi}\|^2},
\end{align}
where $\beta \leq 1$ is a small positive number (typically ranging from $0.01$ to $1$ m$^{-2}$ \cite{3gpp2020channel}) to reflect the density of obstacles in the propagation environment. A larger $\beta$ indicates a denser obstacle environment and hence a lower probability of maintaining a clear LoS path.
Accordingly, the instantaneous channel coefficient $h$ in the presence of LoS blockage can be expressed as\footnote{The non-line-of-sight (NLoS) channel component is not considered in the channel model, as NLoS propagation at millimeter-wave and higher frequency bands typically suffers from severe attenuation, making its contribution negligible in future high-frequency communication systems \cite{ding2025los,rappaport2017overview,xiao2017millimeter}.}
\begin{align}
h = \gamma h^{\text{LoS}}.
\end{align}

\subsection{Problem Formulation}
In this work, we aim to maximize the average received SNR at the user, subject to antenna placement constraints. To begin, we express the instantaneous received SNR as
\begin{align}
\text{SNR}(\tilde x) =\rho |h|^2,
\end{align}
where $\rho = P/\sigma^2$, with $P$ denoting the transmit power and $\sigma^2$ representing the noise power at the receiver.
To analyze the performance, we evaluate the expectation of the received SNR by averaging over both the LoS blockage indicator $\gamma$:
\begin{subequations}
    \begin{align}
        \mathbb{E}[|h|^2] &= \mathbb{E}_{\gamma}\left[|\gamma h^{\text{LoS}}|^2\right] \label{eqn: eh_expansion 1} \\
        &= \mathbb{E} [\gamma] |h^{\text{LoS}}|^2 \label{eqn: eh_expansion 4}\\
        &= \frac{\eta e^{-\beta[(\tilde x - \bar x)^2 + C]}}{[(\tilde x - \bar x)^2 + C] e^{2\alpha\tilde x}}, \label{eqn: eh_expansion 5}
    \end{align}
\end{subequations}
where $C = \bar y^2 + d_v^2$, \eqref{eqn: eh_expansion 4} results from the fact that $\mathbb{E}[\gamma] = \mathbb{E}[\gamma^2]$ since $\gamma \in \{0,1\}$, and \eqref{eqn: eh_expansion 5} is obtained by substituting coordinates and applying the LoS probability defined in \eqref{eqn: los probability}.

Therefore, the average SNR maximization problem is formulated as
\begin{subequations} \label{p: snr maximization}
\begin{align}
\max_{\tilde x} ~& \frac{\rho\eta e^{-\beta[(\tilde x - \bar x)^2 + C]}}{[(\tilde x - \bar x)^2 + C] e^{2\alpha\tilde x}}  \\
\st ~& 0 \leq \tilde x \leq x_{\max}, \label{eqn: antenna placement}
\end{align}
\end{subequations}
where the constraint \eqref{eqn: antenna placement} captures the feasible region for antenna placement. The formulated problem reflects the intricate interplay among multiple factors: the pinching-antenna position jointly affects the LoS probability, in-waveguide attenuation, and free-space path loss, each critically affecting the optimal solution.

\subsection{Optimal Pinching-Antenna Position Derivation and Analysis}

We now derive a closed-form pinching-antenna position, explicitly accounting for both in-waveguide attenuation and LoS blockage effects, that maximizes the average received SNR,.
We begin by defining the following function:
\begin{align}
    g(\tilde x) &\triangleq - \ln \left[ \frac{ e^{-\beta[(\tilde x - \bar x)^2 + C]}}{[(\tilde x - \bar x)^2 + C] e^{2\alpha\tilde x}} \right] \notag\\
    &= \beta(\delta^2 + C) + \ln(\delta^2 + C) + 2\alpha \tilde x,
\end{align}
where $\delta \triangleq \bar x - \tilde x$. According to the analysis in \cite{xu2025pinching}, the optimal pinching-antenna position satisfies $\tilde x^* \leq \bar x$, thus $\delta \geq 0$.

Thus, maximizing the average SNR is equivalent to minimizing the function $g(\tilde x)$ within the feasible region
\begin{align} \label{eqn: antenna placement 2}
\min_{0 \leq \tilde x \leq x_{\max}} \beta(\delta^2 + C) + \ln(\delta^2 + C) + 2\alpha \tilde x.
\end{align}

Before deriving the optimal pinching-antenna position, we present the following lemma.
\begin{lem} \label{lem: convexity}
    Suppose that $\beta d_v^2 \geq 1$ holds. Then $g(\tilde x)$ is a strictly convex function of $\tilde x$ over the feasible region.
\end{lem}

\textit{Proof:} See Appendix \ref{appd: convexity}.  \hfill $\blacksquare$

We note that condition $\beta d_v^2 \geq 1$ can be met in practice. For instance, with a typical value $\beta = 0.1$ m$^{-2}$, it is satisfied when $d_v \geq 3.17$ m. 
Under this condition, the convexity of $g(\tilde x)$ ensures that any stationary point is the unique global minimizer within the feasible set. Therefore, the optimal pinching-antenna position can be efficiently obtained by using a numerical method, e.g., the bisection search method. However, to gain further analytical insights and enable a faster performance evaluation, we instead derive a closed-form expression to facilitate deeper analysis. The result is provided below: 

\begin{lem}\label{lemma:closed_form_solution}
Suppose that $\beta d_v^2 \geq 1$ holds. The optimal pinching-antenna position $\tilde{x}^* \in [0, x_{\max}]$ is given by
\begin{align}
    \tilde{x}^* = \left[ \bar{x} - y^* - \frac{\alpha}{3\beta} \right]_0^{x_{\max}},
\end{align}
where $[z]_a^b \triangleq \min\{\max\{z, a\}, b\}$ denotes the projection onto interval $[a, b]$, and $y^* = A + B$ is the unique real root of the depressed cubic equation $y^3 + p y + q = 0$
with coefficients
\begin{align} \label{eqn: pq}
    p = C + \frac{1}{\beta} - \frac{\alpha^2}{3\beta^2},\ \ q = -\frac{2\alpha^3}{27\beta^3} - \frac{\alpha}{3\beta^2} - \frac{4\alpha C}{3\beta},
\end{align}
and parameters
\begin{align}\label{eqn: uAB}
    u = \sqrt{ \left( \frac{q}{2} \right)^2 + \left( \frac{p}{3} \right)^3 }, \
    A = \sqrt[3]{ -\frac{q}{2} + u }, \
    B = \sqrt[3]{ -\frac{q}{2} - u }.
\end{align}
\end{lem}

\textit{Proof:} Define $\delta \triangleq \bar{x} - \tilde{x}$. 
Note that the optimal pinching-antenna position satisfies the first-order optimality condition $g'(\tilde{x}) = 0$. This leads to the following cubic equation in $\delta$:
\begin{align}
    \beta\delta^3 - \alpha\delta^2 + (\beta C + 1)\delta - \alpha C = 0. \label{eq:stationary_cubic_new}
\end{align}
To solve this cubic equation, we first normalize it by dividing both sides by $\beta > 0$:
\begin{align}
    \delta^3 - \frac{\alpha}{\beta}\delta^2 + \left( C + \frac{1}{\beta} \right)\delta - \frac{\alpha C}{\beta} = 0.
\end{align}
The Tschirnhaus substitution \cite{cardano2007rules}, $\delta = y + \frac{\alpha}{3\beta}$, is applied to eliminate the quadratic term, resulting in the standard depressed cubic form:
\begin{align}
    y^3 + p y + q = 0,
\end{align} with coefficients $p$ and $q$ defined in \eqref{eqn: pq}.
These expressions result from expanding and rearranging $\delta^3$ and $\delta^2$ after substituting $\delta = y + \frac{\alpha}{3\beta}$.

Strict convexity of $g(\tilde x)$ guarantees a unique real root for this cubic. Apply Cardano's formula \cite{cardano2007rules} and define $u$, $A$ and $B$ as in \eqref{eqn: uAB}.
The real-valued solution for $y$ is then
\begin{align}
    y^* = A + B.
\end{align}
Returning to the original variable, the optimal offset is $\delta^* = y^* + \frac{\alpha}{3\beta}$, and the optimal pinching-antenna position is given by
\begin{align} \label{eqn: optimal solution}
\tilde x^* = \bar x - \delta^* = \bar x - y^* - \frac{\alpha}{3\beta}.
\end{align}
If this solution falls outside $[0, x_{\max}]$, the minimizer is achieved at the closest boundary.
The proof of the lemma is complete. 
\hfill $\blacksquare$

The optimal closed-form solution provides an efficient way to compute the average data rate. What's more, it also enables deeper analytical insights into how key system parameters, such as LoS blockage density, in-waveguide attenuation level, and waveguide height, influence the optimal antenna placement and system performance.

\subsubsection{Approximation and Analysis}
To obtain deeper insights, we consider an approximate solution of the optimal pinching-antenna position. Specifically, the approximation is based on the assumption that the offset $\delta$ is small, implying that the optimal pinching-antenna position lies close to the user. As will be shown later, this assumption holds well when the in-waveguide attenuation coefficient $\alpha$ is small, which is a condition typically satisfied in practical system settings.
Under this assumption, the first-order condition in \eqref{eq:stationary_cubic_new} can be approximated as
\begin{align}
(\beta C + 1)\delta - \alpha C \approx 0.
\end{align}
This yields
\begin{align}
\delta \approx \frac{\alpha C}{1 + \beta C}.
\end{align}
As seen, for a small $\alpha$, $\delta$ is also a small number. Accordingly, the approximate optimal position is given by
\begin{align} \label{eqn: approximate tilde x}
\tilde x^* \approx \bar x - \frac{\alpha C}{1 + \beta C}.
\end{align}

This expression leads to several intriguing observations:
\begin{itemize}
    \item For fixed $\beta$ and $C$, as $\alpha$ increases, the optimal $\tilde x^*$ moves closer to the feed point to mitigate in-waveguide attenuation;
    \item For fixed $\alpha$ and $C$, increasing $\beta$ shifts $\tilde x^*$ closer to $\bar x$, enhancing the probability of maintaining a LoS link;
    \item For fixed $\alpha$ and $\beta$, smaller $C$ (i.e., shorter user–to-waveguide distance) drives $\tilde x^*$ closer to $\bar x$, indicating that the free-space LoS path loss becomes the dominant factor for the received SNR in this case.
\end{itemize}


Beyond analyzing the effects of in-waveguide attenuation and LoS blockage on the optimal pinching-antenna position, it is also important to verify how these factors impact the average data rate. 
To this end, we compare the average data rates of three antenna placement strategies across varying communication region sizes, considering both dense and sparse blockage environments. 
In all three strategies, the effect of in-waveguide attenuation is incorporated into the data rate calculation. 
The strategies differ in how they determine the antenna position: the first approach directly incorporates attenuation into the optimization process using the exact formulation in Eq. \eqref{eqn: optimal solution}; the second uses an analytical approximation derived in Eq. \eqref{eqn: approximate tilde x} to simplify computation; and the third disregards in-waveguide attenuation during antenna placement optimization and directly places the antenna above the user, i.e., $\tilde x = \bar x$, to minimizing the free-space path loss and LoS blockage probability.

\begin{figure}[!t]
	\centering
	\includegraphics[width=0.88\linewidth]{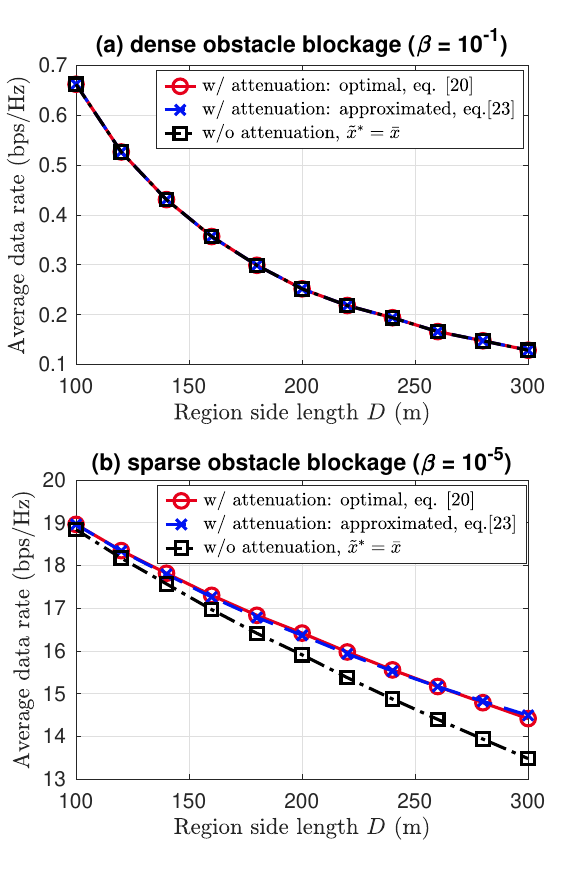}\\
        \captionsetup{justification=justified, singlelinecheck=false, font=small}	
        \caption{Average achievable data rate versus communication region side length $D$ for pinching-antenna system.} \label{fig: data rate c}  \vspace{-3mm}
\end{figure}

Fig. \ref{fig: data rate c} evaluates the average data rates of the schemes versus different communication region side lengths, where the simulation parameters are set as $\alpha = 0.0092$ m$^{-1}$, $f = 28$ GHz, $\sigma^2 = -110$ dBm, $P = 40$ dBm, and $d_v = 10$ m.
As shown in Fig. \ref{fig: data rate c}, in the dense obstacle blockage scenario (e.g., $\beta = 10^{-1}$), the performance gap among the three strategies is minimal. This is because, in such environments, free-space path loss and frequent blockage dominate system performance, making the additional impact of in-waveguide attenuation relatively insignificant. However, in the sparse obstacle blockage scenario (e.g., $\beta = 10^{-5}$), the performance loss due to neglecting in-waveguide attenuation becomes more pronounced as the communication region becomes larger. This is due to the fact that, with a high probability of LoS transmission, both free-space path loss and in-waveguide attenuation jointly limits the achievable rate. These findings highlight the importance to optimize the pinching-antenna position to fully exploit system performance, especially in large coverage areas.
Notably, the data rate obtained using the approximate positioning strategy aligns closely with that of the optimal solution across all scenarios, confirming the analytical approximation's accuracy and practical value for pinching-antenna system design.

\subsection{Analytical Average Data Rate Derivation and Practical System Design Implication}

To further understand the effect of in-waveguide attenuation across different levels of LoS blockage, we analytically investigate the average data rate gap between the following two pinching-antenna placement schemes: one that accounts for in-waveguide attenuation during antenna optimization and one that does not. For analytical tractability, we use the approximated pinching-antenna position from \eqref{eqn: approximate tilde x} for the scheme considering in-waveguide attenuation. By deriving and comparing the average data rates achieved by both approaches, we aim to reveal fundamental performance trends and establish practical guidelines for the deployment and optimization of pinching-antenna systems in realistic environments. The following proposition provides a closed-form approximation of the average data rate degradation incurred when in-waveguide attenuation is ignored during antenna placement.

\begin{prop} \label{prop: avg_rate_loss}
    Let $\Delta R = R_{\mathrm{w}} - R_{\mathrm{wo}}$ denote the data rate gap between antenna placement strategies that respectively consider and ignore in-waveguide attenuation. The expected value of the data rate loss $\mathbb{E}_{\psib}[\Delta R]$ can be approximated by
    \begin{align}
        &\mathbb{E}_{\psib}[\Delta R] \notag\\
        &\approx \frac{\alpha^2}{\beta \ln 2} \left[
            1 \!-\! \frac{2}{D \sqrt{\beta(1 \!+\! \beta d_v^2)}} \arctan \left(
                \frac{\sqrt{\beta}\, D/2}{\sqrt{1 \!+\! \beta d_v^2}}
            \right)
        \right]. \label{eq:avg_rate_loss}
    \end{align}
\end{prop}

\textit{Proof:} See Appendix \ref{appd:avg_rate_loss}. \hfill $\blacksquare$

According the proof for Proposition \ref{prop: avg_rate_loss}, it is noted that \eqref{eq:avg_rate_loss} is tight in the high-SNR regime. This closed-form expression explicitly quantifies the average data rate loss resulting from neglecting in-waveguide attenuation during pinching-antenna placement. Importantly, this result characterizes how the data rate loss depends jointly on key system parameters, including the in-waveguide attenuation coefficient $\alpha$, the blockage density parameter $\beta$, the waveguide height $d_v$, and the communication region size $D$. Building on this expression, several important insights can be drawn regarding the practical significance of in-waveguide attenuation in pinching-antenna system design.

\begin{coro} \label{coro: small beta}
    In the sparse obstacle blockage scenario, i.e., $\beta \to 0$, the expected value of the data rate loss $\mathbb{E}_{\psib}[\Delta R]$ reduces to:
    \begin{align} \label{eqn: rate loss sparse blockage}
        \lim_{\beta \to 0} \mathbb{E}_{\psib}[\Delta R] = \frac{\alpha^2}{\ln 2} \left(\frac{D^2}{12}  + d_v^2\right).
    \end{align}
\end{coro}

\textit{Proof:} See Appendix \ref{appd: small beta}. \hfill $\blacksquare$

The result in \eqref{eqn: rate loss sparse blockage} perfectly matches the expression given in \cite[Proposition 1]{xu2025pinching}. 
This result explicitly shows that the data rate loss linearly with both the square of the in-waveguide attenuation coefficient and the square of the communication region's side length.
Corollary \ref{coro: small beta} also reveals that in environments characterized by sparse obstacle blockage, where LoS paths are almost always available,the received signal strength is determined jointly by both in-waveguide attenuation and free-space path loss. 
Moreover, as the region grows, the negative effect of ignoring in-waveguide attenuation becomes more prominent. This observation suggests an important practical guideline: in scenarios with sparse obstacle blockage, it is advantageous to deploy multiple shorter waveguides to cover large areas, thereby reducing the cumulative effect of in-waveguide attenuation.

\begin{coro} \label{coro: large D}
    For a large communication area with fixed $\beta$, such that $\beta D^2 \gg 1$, the expected value of the data rate loss $\mathbb{E}_{\psib}[\Delta R]$ admits the following asymptotic form:
    \begin{align}
        \lim_{D \to \infty} \mathbb{E}_{\psib}[\Delta R]
        = \frac{\alpha^2}{\beta \ln 2}.
    \end{align}
\end{coro}

\textit{Proof:} See Appendix \ref{appd: large D}. \hfill $\blacksquare$

Corollary \ref{coro: large D} shows that as the communication region becomes very large (i.e., $\beta D^2 \gg 1$), the average data rate loss incurred by ignoring in-waveguide attenuation approaches a constant value, $\frac{\alpha^2}{\beta \ln 2}$, which depends solely on the in-waveguide attenuation coefficient and the LoS blockage parameter. For typical parameter values, such as $\alpha = 0.0092$ m$^{-1}$ and $\beta = 0.1$ m$^{-2}$, the value of this constant loss is just $0.0012$ bps/Hz. This indicates that, even in very large and blockage-dense regions, the performance loss caused by neglecting in-waveguide attenuation is practically negligible.

Collectively, these results offer clear design guidelines for pinching-antenna systems under different blockage conditions: 
\begin{itemize}
    \item In environments with sparse obstacle blockage, our analysis suggests that the average data rate loss caused by in-waveguide attenuation increases with both the in-waveguide attenuation and the size of the coverage area. To mitigate this performance loss in large communication regions, it is preferable to deploy multiple shorter waveguides rather than relying on a single long one. This approach ensures that in-waveguide attenuation remains limited, thereby preserving high system performance.
    \item In contrast, for dense obstacle blockage scenarios, our results reveal that the performance loss incurred by neglecting in-waveguide attenuation becomes negligible, even as the communication area grows large. This is because, in such environments, the free-space path loss and LoS blockage dominate the overall signal attenuation, making the effect of in-waveguide loss much less significant.
\end{itemize}
In summary, the pinching-antenna system design should carefully consider the blockage characteristics of the deployment environment. For sparse obstacle blockage environment, waveguide segmentation is recommended to combat attenuation effects, while for dense obstacle blockage environment, the impact of in-waveguide attenuation can often be safely ignored in practical system design.

\section{Pinching-Antenna System Design in the Multi-User Case} \label{sec: multiple user}

This section considers a downlink transmission scenario where $N$ pinching antennas are distributed across $N$ independent waveguides to simultaneously serve $M$ single-antenna users, with $M \leq N$. Each waveguide is aligned along the $x$-axis at a height $d_v$ and arranged uniformly across the coverage region. Let $d_h$ denote the spacing between adjacent waveguides, which is assumed to satisfy $d_h = \frac{D}{N-1} \gg \lambda$ to avoid inter-waveguide interference.
The coordinate of the feed point on the $n$-th waveguide is given by
$\boldsymbol{\psi}_{0,n} = [0, \left(n-1\right)d_h - \frac{D}{2}, d_v ], \forall n \in \mathcal{N} \triangleq \{1,2,\dots,N\}$. 
The position of the pinching antenna on the $n$-th waveguide is denoted by $\boldsymbol{\psi}^{\mathrm{Pin}}_n = [\tilde{x}_n, \left(n-1\right)d_h - \frac{D}{2}, d_v]$, and the location of user $m$ is denoted by $\boldsymbol{\psi}_m = [\bar{x}_m, \bar{y}_m, 0], \forall m \in \mathcal{M} \triangleq \{1,2,\dots,M\}$.

Based on the observation in the single-user case that the effect of in-waveguide attenuation becomes negligible under typical LoS blockage, we omit this factor in the multi-user analysis for simplicity. Under this assumption, the LoS channel between user $m$ and the $n$-th pinching antenna is modeled as
\begin{align} \label{eqn: los channel}
    h_{m,n}^{\text{LoS}} =\frac{\eta^{\frac{1}{2}}\,e^{-j \left(\frac{2\pi}{\lambda} \sqrt{(\tilde x_n - \bar x_m)^2 + C_{m,n}} + \frac{2\pi}{\lambda_g} \tilde x_n \right)}}{\left[(\tilde x_n - \bar x_m)^2 +  C_{m,n}\right]^{\frac{1}{2}}},
\end{align}
where $C_{m,n} = (\bar y_m - \tilde y_n)^2 + d_v^2$ with $\tilde y_n = (n-1)d_h - \frac{D}{2}$. 
For user $m$ and pinching antenna $n$, the effective channel between them is given by
\begin{align}
    h_{m,n} = \gamma_{m,n} h_{m,n}^{\text{LoS}},
\end{align}
where $\gamma_{m,n} \in \{0,1\}$ denotes the LoS blockage indicator between antenna $n$ and user $m$. The probability of having a LoS path between the $n$-th antenna and user $m$ follows the distance-based model:
\begin{align}
    \mathbb{P}(\gamma_{m,n} = 1) = e^{-\beta \left[ (\tilde x_n - \bar x_m)^2 + C_{m,n} \right]}.
\end{align}

Let $\hb_m = [h_{m,1}, ..., h_{m,N}]^\top$ denote the overall channel vector from all antennas to user $m$. The received signal at user $m$ is
\begin{align}
    y_m = \hb_m^\top \vb_m s_m + \sum_{i \neq m} \hb_m^\top \vb_i s_i + n_m,
\end{align}
where $\mathbf{v}_m \in \mathbb{C}^N$ is the beamformer assigned to user $m$, $s_m$ is the transmitted symbol, and $n_m \sim \mathcal{CN}(0, \sigma_m^2)$ represents AWGN.

Our objective is to jointly design $\{\tilde{x}_n\}_{n=1}^N$ and $\{\mathbf{v}_m\}_{m=1}^M$ to maximize the average sum rate, under random LoS blockages. Let $\tilde{\mathbf{x}} = [\tilde{x}_1, \ldots, \tilde{x}_N]^\top$ and $\mathbf{V} = [\mathbf{v}_1, \ldots, \mathbf{v}_M]$. The optimization problem is given by
\begin{subequations} \label{p: ergodic_sum_rate_maximization}
    \begin{align}
        \max_{\tilde \xb,\, \Vb} ~&\mathbb{E}_{\{\gamma_{m,n}\}} \left[
        \sum_{m=1}^M \log_2\left(1 + \frac{|\hb_m^\top \vb_m|^2}{\sum_{i \neq m}|\hb_m^\top \vb_i|^2 + \sigma_m^2}\right)
        \right] \label{eqn: average sum rate max}\\
        \st~ & \sum_{m=1}^M \|\vb_{m}\|^2 \leq P_{\max}, \\
        & 0 \leq \tilde x_n \leq x_{\max}, \quad \forall n \in \mathcal{N},
    \end{align}
\end{subequations}
where the expectation $\mathbb{E}_{\{\gamma_{m,n}\}}$ is taken over all realizations of the LoS blockage indicators $\gamma_{m,n}$. 

Problem \eqref{p: ergodic_sum_rate_maximization} is particularly challenging to solve for the following two reasons. First, the objective function \eqref{eqn: average sum rate max} requires averaging over random LoS blockage indicators, which introduces significant stochasticity and necessitates integration over their distributions. Second, the beamforming vectors and antenna positions are intricately coupled in the sum-rate expression. This complex coupling, combined with the randomness, results in a highly nonconvex optimization landscape with many local optima, making the problem extremely difficult to solve using conventional optimization techniques. To handle the stochasticity and nonconvexity of problem \eqref{p: ergodic_sum_rate_maximization}, a dynamic SAA-based algorithm is devised in this work.

\subsection{Proposed Dynamic SAA-Based Algorithm}
The SAA method has been used for handling randomness in stochastic optimization problems. It approximates the expectation in the objective function by computing the empirical average over a finite number of randomly generated samples, thereby transforming the original stochastic problem into a more tractable deterministic one \cite{kleywegt2002sample}. However, the conventional SAA method cannot be directly applied to our problem because the distributions of the random variables—particularly the LoS blockage indicators—depend explicitly on the optimization variables, particularly the pinching-antenna positions.

To address this challenge, we propose a dynamic SAA method, in which, at each iteration, the random samples are regenerated based on the current antenna positions. This dynamic resampling strategy allows the algorithm to accurately reflect the evolving system configuration and appropriately capture the impact of antenna placement on the LoS blockage probabilities\footnote{This dynamic sampling strategy is conceptually related to batch stochastic approximation and online optimization techniques, which have been shown to improve convergence and robustness in problems with evolving stochastic components \cite{nemirovski2009robust,shapiro2021lectures}.}. 
Following this idea, the algorithm proceeds in the following iterative manner. First, given a set of pinching-antenna positions, we randomly generate multiple independent channel realizations. Using these samples, an empirical average sum-rate maximization problem is constructed and solved to update both the beamforming vectors and the pinching-antenna positions. Once updated, new channel samples are generated based on the current geometry, and the procedure is repeated until convergence. We now present the details of the proposed dynamic SAA-based algorithm.

Let $L$ denote the number of independent random channel realizations drawn at each iteration, where each realization $\ell = 1, \dots, L$ includes a set of LoS blockage indicators $\{\gamma_{m,n}^{(\ell)}\}$. For a given antenna position vector $\tilde{\xb}$, we generate each realization according to:
\begin{align}
    &\gamma_{m,n}^{(\ell)} \sim \mathrm{Bernoulli}\left(e^{-\beta\left[(\tilde{x}_n - \bar x_m)^2 + C_{m,n}\right]}\right), \ \ \forall m \in \Mset, n \in \Nset.
\end{align}
Then, the original stochastic problem \eqref{p: ergodic_sum_rate_maximization} can be approximated by the deterministic problem:
\begin{subequations}\label{eq:saa_dynamic}
\begin{align}
    \max_{\tilde{\xb},\, \Vb}~ &\frac{1}{L}\sum_{\ell=1}^{L}\sum_{m=1}^{M}\log_2\left(1+\frac{|\hb_m^{(\ell)\top}\vb_m|^2}{\sum_{i\neq m}|\hb_m^{(\ell)\top}\vb_i|^2+\sigma_m^2}\right)\label{eq:saa_dynamic_obj}\\
    \text{s.t.}~~ &\sum_{m=1}^{M}\|\vb_m\|^2\leq P_{\max},\\
    &0\leq\tilde{x}_n\leq x_{\max},\quad\forall n\in\mathcal{N},
\end{align}
\end{subequations}
where $\hb_m^{(\ell)}$ denotes the channel vector for user $m$ under the $\ell$-th channel realization, which is computed based on the sampled LoS blockage indicators $\{\gamma_{m,n}^{(\ell)}\}$ for all $n$. Specifically, each element of $\hb_m^{(\ell)}$ is given by
\begin{align} \label{eqn: composite channel}
    h_{m,n}^{(\ell)} = \gamma_{m,n}^{(\ell)} h_{m,n}^{\mathrm{LoS}}(\tilde x_n),
\end{align}
where $h_{m,n}^{\mathrm{LoS}}(\tilde x_n)$ is given by \eqref{eqn: los channel}, which is the deterministic LoS channel determined by pinching-antenna position $\tilde x_n$ .

Compared to the original stochastic formulation in problem \eqref{p: ergodic_sum_rate_maximization},  problem \eqref{eq:saa_dynamic} is deterministic for a given set of sampled channel realizations, making it more tractable. Notably, the constraints involving the pinching-antenna positions and the beamforming vectors are decoupled, which naturally lends itself to an alternating optimization (AO)  framework. The algorithm iteratively updates the beamforming vectors and pinching-antenna positions. Specifically, at the $t$-th iteration, the following two steps are performed:

\subsubsection{Beamforming Vector Optimization}
With the antenna positions $\tilde{\xb}^{(t)}$ fixed, we solve the following subproblem:
\begin{subequations} \label{p: beamformer wmmse}
    \begin{align}
        \max_{\Vb}~ &\frac{1}{L}\sum_{\ell=1}^{L}\sum_{m=1}^{M}\log_2\left(1+\frac{|\big(\hb_m^{(\ell)}(\tilde{\xb}^{(t)})\big)^\top\vb_m|^2}{\sum_{i\neq m}|\big(\hb_m^{(\ell)}(\tilde{\xb}^{(t)})\big)^\top\vb_i|^2+\sigma_m^2}\right) \label{eqn: beamform}\\
        \text{s.t.}~ &\sum_{m=1}^{M}\|\vb_m\|^2\leq P_{\max}.
    \end{align}
\end{subequations}
This subproblem is a sum-rate maximization problem over a finite set of sampled channels, which, although nonconvex, can be efficiently solved using the WMMSE algorithm. 
Specifically, by applying WMMSE method \cite{shi2011iteratively}, the problem to be solved is given by
\begin{align}
    \min_{\{u_m^{(\ell)}\}, \{w_m^{(\ell)}\}, \Vb}~ 
    &\frac{1}{L}\sum_{\ell=1}^L \sum_{m=1}^M \left( w_m^{(\ell)} e_m^{(\ell)} - \log w_m^{(\ell)} \right) \\
    \text{s.t.}~ & \sum_{m=1}^M \|\vb_m\|^2 \leq P_{\max},
\end{align}
where $u_m^{(\ell)}$ is the receiver coefficient, $w_m^{(\ell)}$ is the minimum mean error (MSE) weight, and $e_m^{(\ell)}$ is the MSE for user $m$ in realization $\ell$. These variables can be iteratively updated. In particular, given fixed beamformers $\Vb$, the optimal receiver and weight updates are given in closed form:
\begin{subequations}
    \begin{align}
        u_m^{(\ell)} &= \frac{\big(\hb_m^{(\ell)}(\tilde x^{(t)})\big)^\top \vb_m}{\sum_{i=1}^M \big|\big(\hb_m^{(\ell)}(\tilde x^{(t)})\big)^\top \vb_i\big|^2 + \sigma_m^2}, \\
        e_m^{(\ell)} &= 1 - 2 \mathcal{R} \left\{ u_m^{(\ell)*} \big(\hb_m^{(\ell)}(\tilde x^{(t)})\big)^\top \vb_m \right\} \notag\\
        &\ \ \ \ + |u_m^{(\ell)}|^2 \left(\sum_{i=1}^M \big|\big(\hb_m^{(\ell)}(\tilde x^{(t)})\big)^\top \vb_i\big|^2 + \sigma_m^2 \right), \\
        w_m^{(\ell)}& = \left(e_m^{(\ell)}\right)^{-1}.
    \end{align}
\end{subequations}
Then, with the receivers and weights fixed, the optimal beamforming vectors $\Vb$ are obtained by solving the following convex quadratic program:
\begin{subequations}
    \begin{align}
        \min_{\Vb}~ &\frac{1}{L}\sum_{\ell=1}^L \sum_{m=1}^M w_m^{(\ell)} e_m^{(\ell)} \\
        \text{s.t.}~ & \sum_{m=1}^M \|\vb_m\|^2 \leq P_{\max}.
    \end{align}
\end{subequations}  
This can be efficiently solved via standard convex optimization tools, or, in closed-form using Lagrangian duality \cite{shi2011iteratively}. The above optimization steps are repeated until convergence.

\subsubsection{Antenna-Position Update}
With beamformers $\Vb^{(t+1)}$ fixed, we update the antenna positions by solving:
\begin{subequations} \label{eqn: position update}
    \begin{align}
        \max_{\tilde{\xb}}~ &\frac{1}{L}\sum_{\ell=1}^{L}\sum_{m=1}^{M}\log_2\left(1+\frac{\big|\big(\hb_m^{(\ell)}(\tilde{\xb})\big)^\top\vb_m^{(t+1)}\big|^2}{\sum_{i\neq m}\big|\big(\hb_m^{(\ell)}(\tilde{\xb})\big)^\top\vb_i^{(t+1)}\big|^2+\sigma_m^2}\right) \\
        \text{s.t.}~ &0\leq\tilde{x}_n\leq x_{\max},~\forall n\in\mathcal{N}.
    \end{align}  
\end{subequations}

By exploiting the inherent decomposable structure of the optimization problem, the position of each pinching antenna can be optimized individually, while the other antennas fixed. However, each resulting one-dimensional subproblem remains highly challenging due to the intricate dependency of the channel phase shifts---both in free space and within the waveguide---on antenna positions at the scale of the carrier wavelength. This results in a highly non-convex objective function characterized by rapid oscillations and hence numerous local optima. Standard gradient-based optimization algorithms are often ineffective under such conditions, as they are prone to getting trapped in suboptimal solutions.
To address this issue, we adopt the particle swarm optimization (PSO) method to sequentially optimize each pinching-antenna position. The detailed procedure of the proposed PSO-based antenna position optimization is provided as follows.

\paragraph*{A) Swarm Initialization}
We randomly initialize a swarm comprising $S$ particles, each associated with a candidate solution for the pinching-antenna position and an initial velocity. Specifically, the position $x_n^{(s)}(0)$ and velocity $v_n^{(s)}(0)$ for the $s$-th particle corresponding to the $n$-th antenna are initialized according to:
\begin{subequations}
    \begin{align}
          x_n^{(s)}(0) &\sim \mathcal U(0,x_{\max}), \ \ \forall s \in \mathcal{S}, \\
          v_n^{(s)}(0) &\sim \mathcal U\bigl(-0.2x_{\max},0.2x_{\max}\bigr)\ \ \forall s \in \mathcal{S},
    \end{align}
\end{subequations}
where $\mathcal{S} \triangleq \{1,...,S\}$.
Then, we evaluate the initial fitness of each particle based on the objective function $F_n(\tilde{x}_n)$, which is defined as
\begin{align}
    &F_n(\tilde{x}_n) = \notag\\
    &\frac{1}{L}\sum_{\ell=1}^{L}\sum_{m=1}^{M}\log_2\left(1+\frac{\big|\big(\hb_m^{(\ell)}(\tilde{x}_n)\big)^\top\vb_m^{(t+1)}\big|^2}{\sum_{i\neq m}\big|\big(\hb_m^{(\ell)}(\tilde{x}_n)\big)^\top\vb_i^{(t+1)}\big|^2+\sigma^2}\right).
\end{align}
Based on these initial fitness values $\mathcal R^{(s)}=F_n\bigl(x_n^{(s)}(0)\bigr), \forall s \in \mathcal{S}$, the personal best positions of each particle are initialized as $p_n^{(s)}=x_n^{(s)}(0)$, and the global best position is set to 
\begin{align}
g_n(0)=\arg\max_{s=1,\dots,S}\mathcal R^{(s)}.
\end{align}

\paragraph*{B) Particle Evolution}
The particle swarm is iteratively updated through successive evolutions. In each iteration $k=0,1,\dots,K_{\max}-1$, the velocity and position of the $s$-th particle are updated according to the following equations:
\begin{subequations}
\begin{align}
  v_n^{(s)}(k+1)
  &= w(k)\,v_n^{(s)}(k)
     + c_1 r_1\bigl[p_n^{(s)}-x_n^{(s)}(k)\bigr] \notag\\
     &\quad + c_2 r_2\bigl[g_n(k)-x_n^{(s)}(k)\bigr],\\
  x_n^{(s)}(k+1) &= \mathrm{clip} \bigl(x_n^{(s)}(k)+v_n^{(s)}(k+1),\,0,\,x_{\max}\bigr),
\end{align}
\end{subequations}
where $\mathrm{clip}(\cdot,a,b)$ denotes the element-wise projection onto the interval $[a, b]$ to ensure that each updated particle position remains within the feasible range, $r_1,r_2\sim\mathcal U(0,1)$ are independent random variables introducing stochasticity, and the inertia weight $w(k)$ linearly decreases as $w(t)=0.9-0.5\,k/K_{\max}$. The cognitive and social acceleration coefficients are set to $c_1=c_2=2$ to realize a balance between the exploration and exploitation.
After updating the positions, the fitness of each particle is recalculated. The personal best $p_n^{(s)}$ for each particle is updated if an improved fitness value is found. Likewise, the global best position $g_n(k+1)$ is refreshed if a particle surpasses the previous global best fitness. The PSO algorithm continues until a predefined convergence criterion is satisfied.

\begin{algorithm}[t]  \small
    \caption{Dynamic SAA-Based Joint Beamforming and Antenna Position Optimization}
    \label{alg: dynamic saa}
    \begin{algorithmic}[1]
        \STATE \textbf{Input:} Number of samples $L$, max iterations $T_{\max}$, initial positions $\tilde{\xb}^{(0)}$
        \FOR{$t = 0, 1, \dots, T_{\max}$}
            \STATE Generate $L$ independent samples $\{ \gamma_{m,n}^{(\ell)}\}_{\ell=1}^L$ based on $\tilde{\xb}^{(t)}$
            \STATE \textbf{Beamformer Update:} With $\tilde{\xb}^{(t)}$ and the $L$ channel realizations fixed, update $\Vb^{(t+1)}$ by solving Problem \eqref{p: beamformer wmmse} via WMMSE method.
            \STATE \textbf{Antenna Position Update:} With $\Vb^{(t+1)}$ fixed, update each $\tilde{x}_n^{(t+1)}$ sequentially by running the PSO-based algorithm for each pinching antenna.
            \STATE \textbf{Convergence Check:} If stopping criterion is met, \textbf{break}.
        \ENDFOR
        \STATE \textbf{Output:} Optimized antenna positions $\tilde{\xb}^*$ and beamformers $\Vb^*$
    \end{algorithmic}
\end{algorithm}

After updating the pinching-antenna positions to $\tilde{\xb}^{(t+1)}$, the distributions of the LoS blockage indicators, and consequently, the composite channel in \eqref{eqn: composite channel} are changed. To ensure that the empirical average in the optimization accurately represents the true system configuration, we regenerate $L$ fresh samples of the LoS blockage indicators according to the updated antenna positions. This dynamic resampling guarantees that each iteration of the algorithm employs random realizations consistent with the current geometry, thus maintaining statistical correctness throughout the optimization. The proposed dynamic SAA-based algorithm alternately updates the beamformers and pinching-antenna positions, and regenerates the random channel samples after each iteration, until convergence. The complete procedure is summarized in Algorithm \ref{alg: dynamic saa}.

\begin{figure*}[!t]
	\centering
	\includegraphics[width=0.99\linewidth]{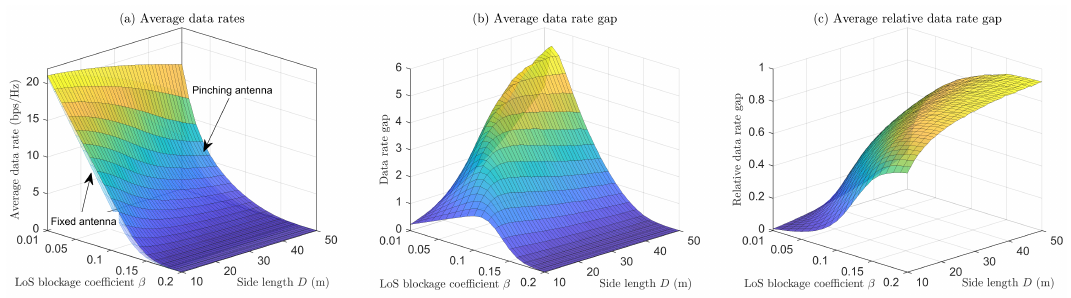}\\
        \captionsetup{justification=justified, singlelinecheck=false, font=small}	
        \caption{\small Average achievable data rate comparison of pinching versus fixed antenna in LoS blockage environments.} \label{fig: data rate 3d}  
\end{figure*} 
 
\section{Simulation Results} \label{sec: simulation}

In this section, numerical experiments are conducted to validate the effectiveness of the proposed pinching-antenna system and the associated optimization algorithms under random LoS blockage. The simulation is carried out using the following settings. The carrier frequency is set to $f_c = 28$ GHz, with a waveguide effective refractive index $n_{\mathrm{eff}} = 1.4$. The height of the waveguide is $d_v = 10$ m. For the dynamic SAA-based algorithm, we use $L = 100$ random realizations of the LoS blockage variables to estimate the expected performance. The AWGN noise power is fixed at $-110$ dBm. Unless otherwise specified, the values of the LoS blockage parameter $\beta$, the total transmit power $P_{\max}$, the coverage area side length $D$, the number of served users $M$, and the number of pinching antennas $N$ are provided in the respective figure captions. For all iterative schemes, convergence is determined based on the variation of the objective value. Specifically, an algorithm is terminated when the relative change between two successive iterations drops below $10^{-3}$, or the number of iterations reaches $T_{\max} = 20$.

We first compares the average data rates achieved by the pinching antenna scheme and the fixed antenna scheme in the single-user pinching-antenna system, under varying levels of LoS blockage and communication region sizes, as shown in Fig. \ref{fig: data rate 3d}. The LoS blockage coefficient $\beta$ varies from $0.01$ to $0.2$, and the region side length $D$ ranges from $10$ to $50$ meters. For the pinching antenna scheme, the antenna is adaptively positioned to coincide with the user's $x$-coordinate ($\tilde{x} = \bar{x}$), whereas in the fixed antenna scheme, the antenna position remains static at position $[\frac{D}{2}, 0, d_v]$. The average data rates are computed by averaging over multiple random user locations for each configuration.

Fig. \ref{fig: data rate 3d}(a) shows the average achievable data rates of both schemes as functions of the LoS blockage coefficient $\beta$ and the region side length $D$. The surface corresponding to the pinching antenna scheme consistently lies above that of the fixed antenna scheme, demonstrating the performance gain achievable through dynamic antenna positioning. Both schemes exhibit a monotonic decrease in data rate as either $\beta$ or $D$ increases, reflecting the adverse impact of increased blockage probability and larger communication areas.
Fig. \ref{fig: data rate 3d}(b) illustrates the average data rate gap between the pinching and fixed antenna schemes across varying LoS blockage coefficients and region side lengths. Several key interesting observations can be drawn from Fig. \ref{fig: data rate 3d}(b).
With a fixed LoS blockage coefficient $\beta$, the rate gap increases with the side length $D$. This demonstrates that, as the communication region expands, the advantage of adaptively positioning the pinching antenna becomes more significant, as the combined effects of large-scale fading and increased probability of LoS blockage can be effectively mitigated by placing the antenna closer to the user.
Conversely, for a fixed small $D$ (e.g., $D = 10$ m), the rate gap exhibits a non-monotonic relationship with respect to $\beta$. Specifically, as $\beta$ increases from very small values, the rate gap first grows, reflecting that modest blockage increases the likelihood of losing the LoS path for the fixed antenna, while the pinching antenna can still adapt to maintain a favorable channel. However, as $\beta$ continues to rise, indicating very dense obstacle blockage, the overall performance of both schemes becomes limited by the low probability of LoS, causing the rate gap to diminish.
In scenarios with very low blockage (small $\beta$) or a small communication region ($D$), both the pinching and fixed antenna schemes achieve high data rates, resulting in only minor performance differences. Conversely, under severe blockage conditions (large $\beta$) or for very large coverage areas ($D$), LoS links are frequently blocked for both schemes, which diminishes the relative advantage of dynamic antenna positioning.
Fig. \ref{fig: data rate 3d}(c) presents the average relative data rate gap between the pinching and fixed antenna schemes, defined as $(R_{\text{Pin}} - R_{\text{Fix}})/R_{\text{Pin}}$, where $R_{\text{Pin}}$ and $R_{\text{Fix}}$ denote the average data rates achieved by the pinching and fixed antenna schemes, respectively. As shown in Fig. \ref{fig: data rate 3d}(c), the relative gap is negligible when the obstacle is sparse or the communication region is small. However, as either the blockage coefficient $\beta$ or the region size $D$ increases, the relative gap grows significantly. 
Overall, these results confirm that dynamic pinching-antenna positioning can significantly enhance the achievable data rates in practical LoS blockage environments, particularly for moderate to large coverage areas and non-negligible blockage conditions. This highlights the practical value of pinching-antenna systems for robust wireless access in challenging propagation conditions.


 \begin{figure}[!t]
 	\centering
 	\includegraphics[width=0.92\linewidth]{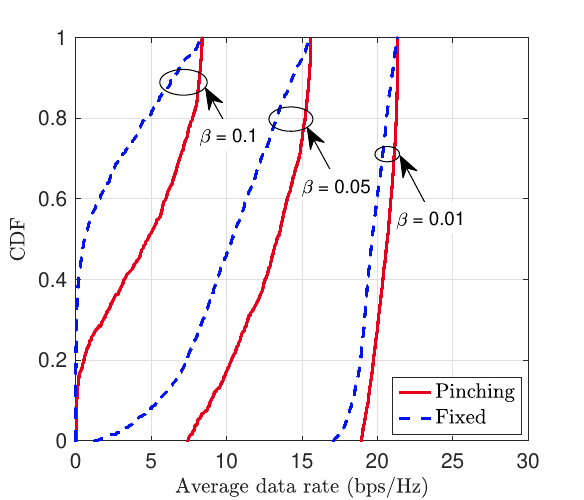}\\
         \captionsetup{justification=justified, singlelinecheck=false, font=small}	
         \caption{CDFs of the achievable data rate for pinching and fixed antenna schemes under different LoS blockage coefficients $\beta$ with side length $D=20$ m.} \label{fig: data rate cdf}  
 \end{figure} 

Fig. \ref{fig: data rate cdf} presents the cumulative distribution functions (CDFs) of the achievable data rates for both pinching and fixed antenna schemes under different LoS blockage coefficients. Each curve is generated from $10^3$ independent random user deployments. As seen from Fig. \ref{fig: data rate cdf}, increasing the blockage coefficient $\beta$ leads to a leftward shift in the CDF curves for both schemes, indicating a reduction in achievable data rate. This improvement is due to the increased difficulty of maintaining a reliable LoS connection. The pinching antenna scheme consistently exhibits superior performance than the fixed antennas for all LoS blockage levels, as indicated by its CDFs being positioned further to the right. This advantage becomes more pronounced with a higher $\beta$, demonstrating that the benefit of adaptive antenna positioning is especially significant in more challenging blockage environments. These results highlight the robustness and practical effectiveness of adaptive pinching-antenna positioning, particularly for improving user rates and link reliability in wireless environments with random LoS blockage.

 \begin{figure}[!t]
 	\centering
 	\includegraphics[width=0.92\linewidth]{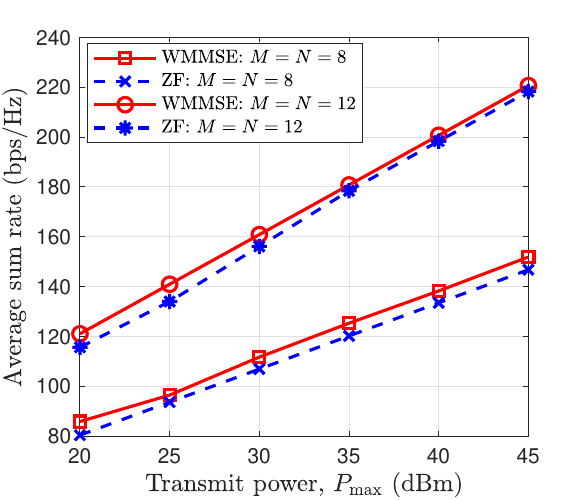}\\
         \captionsetup{justification=justified, singlelinecheck=false, font=small}	
         \caption{Achievable sum rate comparison between the WMMSE-based and ZF-based beamforming schemes in the dynamic SAA framework with $\beta = 0.01$ and $D = 50$ m.} \label{fig: sum rate zf}  
 \end{figure} 
Fig. \ref{fig: sum rate zf} presents a comparison between the proposed WMMSE-based design and the ZF-based baseline, both implemented within the dynamic SAA framework. In the ZF scheme, the transmit beamformers are iteratively updated using zero-forcing precoding, whereas in the WMMSE scheme, the beamforming vectors are optimized via the WMMSE method at each iteration. The WMMSE algorithm is initialized using the ZF beamformer. As shown in Fig. \ref{fig: sum rate zf}, the WMMSE-based design always yields higher sum rates than the ZF-based baseline, although the performance gap remains modest across all tested scenarios. This is because the spatial flexibility of pinching antennas effectively mitigates multiuser interference, allowing even the simpler ZF method to perform competitively. These results indicate that while WMMSE offers performance improvements, ZF remains a practical, low-complexity alternative when computational resources are constrained.

 \begin{figure}[!t]
 	\centering
 	\includegraphics[width=0.92\linewidth]{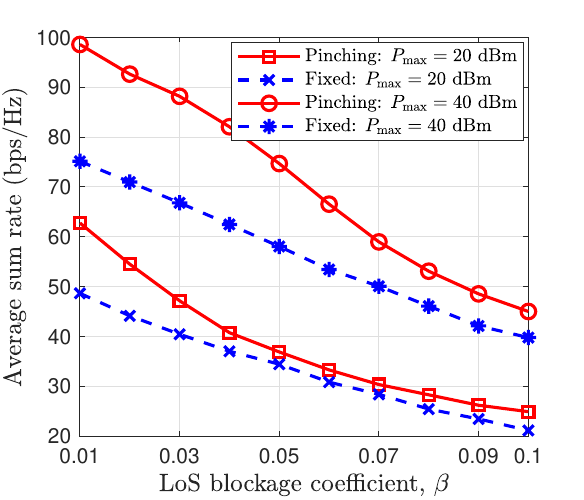}\\
         \captionsetup{justification=justified, singlelinecheck=false, font=small}	
         \caption{Average sum rate versus LoS blockage coefficient $\beta$ for pinching-antenna and fixed-antenna systems with $M=N=4$ and $D = 50$ m.} \label{fig: sum rate beta}  
 \end{figure} 
Fig. \ref{fig: sum rate beta} evaluates the average sum rate of pinching-antenna and fixed-antenna systems under varying levels of LoS blockage, characterized by the blockage coefficient $\beta$. 
As depicted in Fig. \ref{fig: sum rate beta}, the pinching-antenna system always yields a higher data rate than the fixed-antenna baseline across all blockage levels and power settings. This improvement is due to the spatial reconfigurability of pinching antennas which enables the system to dynamically reposition pinching antennas to circumvent blockages and maintain favorable channel conditions. In contrast, fixed antennas lack this adaptability, resulting in degraded performance under severe blockage. These results underscore the robustness of pinching-antenna systems, particularly in maintaining high throughput in dynamic or obstruction-prone environments.

\begin{figure}[!t]
	\centering
	\includegraphics[width=0.92\linewidth]{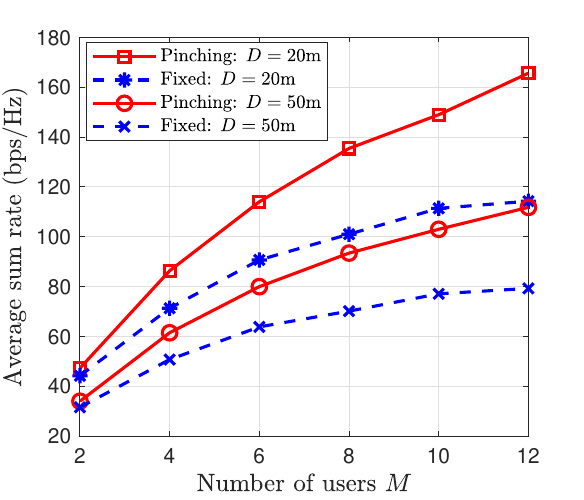}\\
        \captionsetup{justification=justified, singlelinecheck=false, font=small}	
        \caption{Average sum rate versus number of users $M$ for fixed‐ and pinching‐antenna schemes with $\beta = 0.01$ and $N=12$.} \label{fig: sum rate m}  
\end{figure} 
In Fig. \ref{fig: sum rate m}, we investigate how the sum rate scales with the number of users for both systems, under two different deployment areas: $D = 20$ m and $D = 50$ m. For both settings, the pinching-antenna configuration consistently exhibits increasing sum rates as the number of users grows, leveraging multi-user diversity. Notably, the pinching-antenna scheme consistently outperforms the fixed-antenna scheme in both small and large areas, with the performance gap becoming more pronounced as the region size grows. These results confirm the ability of pinching antennas to dynamically reposition allows for better channel conditions, mitigating the impact of path loss and random LoS blockage. In contrast, the fixed-antenna system suffers from degraded performance in larger service areas, as users are more likely to experience unfavorable channel conditions and blockage. These results demonstrate the inherent flexibility of the pinching-antenna architecture, especially in scenarios with an increasing number of users.

\section{Conclusions} \label{sec: conclusion}
This paper explored the design of downlink pinching-antenna systems under practical deployment conditions with random LoS blockage. By integrating in-waveguide attenuation into the system model, we developed a more realistic framework for analyzing system performance in complex environments.
The analysis began with the single-user scenario, where we derived a closed-form expression for the average data rate loss resulting from neglecting in-waveguide attenuation in the antenna placement procedure. This expression revealed how key system parameters, such as waveguide length, attenuation level, and blockage density, jointly affect the system performance, offering useful guidelines for practical pinching-antenna system deployments.
We then extended the investigation to a multi-user MIMO system and formulated a joint optimization problem involving both beamforming and antenna positioning. To account for LoS variability, a dynamic SAA-based approach was proposed, followed by an alternating optimization algorithm leveraging WMMSE and PSO techniques to tackle the resulting stochastic and non-convex problem.
Simulation results validated the effectiveness of the proposed scheme and algorithm, showing that pinching-antenna systems consistently outperform fixed-antenna configurations, especially under high LoS blockage conditions. These findings highlight the potential of spatially reconfigurable antennas in enhancing reliability and throughput in future wireless communication systems.

\begin{appendices}

\section{Proof of Lemma \ref{lem: convexity}} \label{appd: convexity}

First, we derive the first-order derivative of $g(\tilde x)$, which is given by
\begin{align}
    g'(\tilde x) = -2\beta\delta - \frac{2\delta}{\delta^2 + C} + 2\alpha.
\end{align}
where we use the fact that $\frac{\partial \delta}{\partial \tilde x} = -1$. 

To verify the convexity of $g(\tilde x)$, we compute its second-order derivative which is given by
\begin{align}
    g''(\tilde x) = 2\beta + \frac{2}{\delta^2 + C} - \frac{4\delta^2}{(\delta^2 + C)^2}.
\end{align}
Given that $\delta^2 < \delta^2 + C$, we have $\frac{4\delta^2}{(\delta^2 + C)^2} < \frac{4}{\delta^2 + C}$, which leads to the bound
\begin{align}
    g''(\tilde x) > 2\beta - \frac{2}{\delta^2 + C}.
\end{align}
Since $0 \leq \delta \leq \bar x$ and $C \geq d_v^2$, we see that if $\beta d_v^2 \geq 1$, then $g''(\tilde x) \geq 0$ always holds. The proof of the lemma is complete. \hfill $\blacksquare$

\section{Proof of Proposition \ref{prop: avg_rate_loss}} \label{appd:avg_rate_loss}

We first focus on the scheme without in-waveguide attenuation when optimizing the pinching-antenna position. In this case, the pinching antenna is placed at $\tilde x^* = \bar x$, and thus the instantaneous received SNR is given by
\begin{align} \label{eqn: snr_wo}
    \SNR_{\wo} = \frac{\rho \eta e^{-\beta C}}{C e^{2\alpha \bar x} },
\end{align}
where $C = \bar y^2 + d_v^2$.

For the scheme with attenuation when optimizing the pinching-antenna position, we use the approximated solution $\tilde x^* = \bar x - \delta$ with $\delta = \frac{\alpha C}{1 + \beta C}$. The corresponding SNR is given by
\begin{align} \label{eqn: snr_w}
    \SNR_{\w} = \frac{\rho \eta e^{-\beta(\delta^2 + C)}}{(C + \delta^2) e^{2\alpha (\bar x - \delta)}} .
\end{align}

Based on \eqref{eqn: snr_wo} and \eqref{eqn: snr_w}, the instantaneous data rate loss incurred by neglecting the in-waveguide attenuation in the high SNR regime can be approximated as
\begin{align}
    \Delta R 
    &= R_{\w} - R_{\wo} \notag\\
    &\approx \log_2 \left(\frac{\SNR_{\w}}{\SNR_{\wo}} \right) \notag\\
    &= \log_2 \left( \frac{C}{\delta^2 + C} \right) - \frac{\beta}{\ln 2} \delta^2
        + \frac{2\alpha}{\ln 2} \delta. \label{eq:dr-split}
\end{align}

To facilitate analysis, a more tractable form of \eqref{eq:dr-split} is required.  First, considering small $\delta$, we have
\begin{align}
    \log_2 \left( \frac{C}{\delta^2 + C} \right) & \approx \log_2 \left( 1- \frac{\delta^2}{C} \right) \notag\\
    &\approx -\frac{\delta^2}{C \ln 2}.
\end{align}
Therefore, we can rewrite the average data rate as
\begin{align}
    \Delta R \approx -\frac{\delta^2}{C \ln 2} - \frac{\beta \delta^2}{\ln 2} + \frac{2\alpha \delta}{\ln 2} .
\end{align}
Plug $\delta = \frac{\alpha C}{1 + \beta C}$ into the following terms, we have
\begin{align*}
    \delta^2 &= \frac{\alpha^2 C^2}{(1 + \beta C)^2}, \quad
    \frac{\delta^2}{C} = \frac{\alpha^2 C}{(1 + \beta C)^2}, \\
    \beta \delta^2 &= \frac{\beta \alpha^2 C^2}{(1 + \beta C)^2}, \quad
    2\alpha \delta = \frac{2\alpha^2 C}{1 + \beta C}.
\end{align*}
Therefore, we can approximate the instantaneous data rate loss as follows:
\begin{align}
    \Delta R &\approx -\frac{\alpha^2 C}{(1 + \beta C)^2 \ln 2}
                    -\frac{\beta \alpha^2 C^2}{(1 + \beta C)^2 \ln 2} 
                    + \frac{2\alpha^2 C}{(1 + \beta C) \ln 2}\notag\\
            & =   -\frac{\alpha^2 C}{(1 + \beta C) \ln 2}
                    + \frac{2\alpha^2 C}{(1 + \beta C) \ln 2} \notag\\
            & = \frac{\alpha^2 C}{(1 + \beta C) \ln 2}.
\end{align}

Next, the average data rate loss will be focused on. Using the assumption that $\bar{y}$ is uniformly distributed over $[-D/2, D/2]$, we have
\begin{align} 
    \mathbb{E}_{\psib}[\Delta R] &= \frac{1}{D} \int_{-D/2}^{D/2} \frac{\alpha^2 (\bar{y}^2 + d_v^2)}{(1 + \beta (\bar{y}^2 + d_v^2)) \ln 2} d\bar{y} \notag\\
    &= \frac{2\alpha^2}{D \ln 2} \int_0^{D/2} \frac{y^2 + d_v^2}{1 + \beta (y^2 + d_v^2)} d \bar y \notag\\
    &= \frac{2\alpha^2}{D \ln 2} \bigg(\underbrace{\int_0^{D/2} \frac{\bar y^2}{1 + \beta (\bar y^2 + d_v^2)} d \bar y}_{I_1} \notag\\
    & \quad +  d_v^2 \int_0^{D/2} \frac{1}{1 + \beta (\bar y^2 + d_v^2)} d \bar y \bigg). \label{eqn: ER}
\end{align}
Recall that
\begin{align} 
    \frac{\bar y^2}{1 + \beta (\bar y^2 + d_v^2)}
    &= \frac{1}{\beta} - \frac{1 + \beta d_v^2}{\beta [1 + \beta (\bar y^2 + d_v^2)]}.
\end{align}
Therefore, the integral $I_1$ can be writen as
\begin{align} \label{eqn: int 1}
    &\int_0^{D/2} \frac{\bar y^2}{1 + \beta (\bar y^2 + d_v^2)} d\bar y \notag \\
    &= \frac{D}{2\beta} - \frac{1 + \beta d_v^2}{\beta} \int_0^{D/2} \frac{1}{1 + \beta (\bar y^2 + d_v^2)} d \bar y.
\end{align}
Substituting \eqref{eqn: int 1} back to \eqref{eqn: ER}, the average data rate loss can be rewritten as
\begin{align} \label{eqn: ER 2}
    \mathbb{E}_{\psib}[\Delta R] &= \frac{2\alpha^2}{D \ln 2} \left(\frac{D}{2\beta} - \frac{1}{\beta}  \int_0^{D/2} \frac{1}{1 + \beta (\bar y^2 + d_v^2)} d \bar y\right),
\end{align}
Recall that the integral in the above equation can be expressed as follows:
\begin{align} \label{eqn: int 2}
    & \int_0^{D/2} \frac{1}{1 + \beta (\bar y^2 + d_v^2)} d \bar y \notag\\
    & = \frac{1}{\sqrt{\beta (1 + \beta d_v^2)}} \arctan \left( \frac{\sqrt{\beta} D / 2}{\sqrt{1 + \beta d_v^2}} \right), 
\end{align}
which can be used to obtain the following expression for the average data rate loss
\begin{align}
    &\mathbb{E}_{\psib}[\Delta R]\notag\\
    &= \frac{2\alpha^2}{D \ln 2}
        \left[ \frac{D}{2\beta} - \frac{1}{\beta \sqrt{\beta (1 + \beta d_v^2)}} \arctan \left( \frac{\sqrt{\beta} D / 2}{\sqrt{1 + \beta d_v^2}} \right) \right] \notag\\
    &= \frac{\alpha^2}{\beta \ln 2}
        \left[ 1 - \frac{2}{D \sqrt{\beta (1 + \beta d_v^2)}}
            \arctan \left( \frac{\sqrt{\beta} D / 2}{\sqrt{1 + \beta d_v^2}} \right)
        \right]. \label{eqn: average data rate loss 2}
\end{align}
The proof of the proposition is complete. \hfill $\blacksquare$

\section{Proof of Corollary \ref{coro: small beta}} \label{appd: small beta}

In this proof, the case with $\beta \to 0$ will be focused on. Define
\begin{align}
    E = 1 + \beta d_v^2, \qquad z = \frac{\sqrt{\beta} D}{2\sqrt{E}}.
\end{align}
For small $\beta$, the first-order Taylor approximation $(1 + \epsilon)^{-1/2} \approx 1 - \frac{\epsilon}{2}$ can be used to obtain the following approximation:
\begin{align}
    z \approx \frac{\sqrt{\beta} D}{2} \left( 1 - \tfrac{1}{2} \beta d_v^2 \right).
\end{align}
In addition, for small $z$, $\arctan(z)$ can be approximated by
\begin{align}
    \arctan(z) \approx z - \frac{z^3}{3}.
\end{align}
So, we have
\begin{align} \label{eqn: arctan term 1}
    &\arctan \left( \frac{\sqrt{\beta} D}{2\sqrt{E}} \right) \notag \\
    &\approx \frac{\sqrt{\beta} D}{2} \left( 1 - \tfrac{1}{2} \beta d_v^2 \right)
        - \frac{1}{3} \left[ \frac{\sqrt{\beta} D}{2} \left( 1 - \tfrac{1}{2} \beta d_v^2 \right) \right]^3.
\end{align}
Then, the cube term in \eqref{eqn: arctan term 1} can be approximated as follows:
\begin{align} \label{eqn: cube term approximation}
    \left[ \frac{\sqrt{\beta} D}{2} (1 - \tfrac{1}{2} \beta d_v^2) \right]^3  &\overset{(a)}{\approx} \left( \frac{\sqrt{\beta} D}{2} \right)^3 (1 - \tfrac{3}{2} \beta d_v^2 ) \notag \\
    &= \frac{\beta^{3/2} D^3}{8} (1 - \tfrac{3}{2} \beta d_v^2).
\end{align}
where, in (a),  the high-order terms related to the small $\beta$ are dropped.
Therefore, substituting the cube term in \eqref{eqn: arctan term 1} by \eqref{eqn: cube term approximation}, we arrive at
\begin{align} \label{eqn: arctan term approx}
    &\arctan \left( \frac{\sqrt{\beta} D}{2\sqrt{E}} \right) \notag\\
    &\approx \frac{\sqrt{\beta} D}{2} \left(1 - \tfrac{1}{2} \beta d_v^2\right)
    - \frac{1}{3} \frac{\beta^{3/2} D^3}{8} \left(1 - \tfrac{3}{2} \beta d_v^2 \right).
\end{align}

Substituting \eqref{eqn: arctan term approx} into the pre-factor in \eqref{eqn: average data rate loss 2}, $S$ can be expressed as follows:
\begin{align}
    S &\triangleq \frac{2}{D \sqrt{\beta (1 + \beta d_v^2)}} \arctan \left( \frac{\sqrt{\beta} D/2}{\sqrt{1 + \beta d_v^2}} \right) \notag\\
    &\overset{(a)}{\approx} \frac{2\left(1 \!-\! \tfrac{1}{2} \beta d_v^2\right)}{D \sqrt{\beta}}  \bigg[ \frac{\sqrt{\beta} D}{2} \left(1 \!-\! \tfrac{1}{2} \beta d_v^2\right)   - \frac{\beta^{3/2} D^3}{24} \left(1 \!-\! \tfrac{3}{2} \beta d_v^2 \right) \bigg]\notag \\
    &= (1 - \tfrac{1}{2} \beta d_v^2) (1 - \tfrac{1}{2} \beta d_v^2) 
        - \frac{\beta D^2}{12} (1 - \tfrac{3}{2} \beta d_v^2 )(1 - \tfrac{1}{2} \beta d_v^2) \notag\\
    &\overset{(b)}{\approx} (1 - \beta d_v^2 ) - \frac{\beta D^2}{12} \left( 1 - 2 \beta d_v^2 \right),
\end{align}
where, in (a), the approximation $(1 + \epsilon)^{-1/2} \approx 1 - \frac{\epsilon}{2}$ is used for small $\beta$; in (b), the higher-order terms of $\beta$ is dropped.
Therefore, we have
\begin{align}
    1 - S &\approx 1 - (1 - \beta d_v^2) + \frac{\beta D^2}{12} (1 - 2 \beta d_v^2) \notag\\
    & =  \beta d_v^2 + \frac{\beta D^2}{12} + \frac{\beta^2 D^2 d_v^2}{6}.
\end{align}
Finally, by dropping the term related to $\beta^2$ and substituting it back into \eqref{eqn: average data rate loss 2}, the following approximation can be obtained:
\begin{align}
    \mathbb{E}_{\psib}[\Delta R]
    &\approx \frac{\alpha^2}{\beta \ln 2} \left( \beta d_v^2 + \frac{\beta D^2}{12} \right) \notag \\
    &= \frac{\alpha^2}{\ln 2} \left( d_v^2 + \frac{D^2}{12} \right).
\end{align}
The proof of the corollary is complete. \hfill $\blacksquare$

\section{Proof of Corollary \ref{coro: large D}} \label{appd: large D}
In this proof, the case with $D \to \infty$ is focused on.
Define $E = 1 + \beta d_v^2$ and $s_{\max} = \frac{\sqrt{\beta} D}{2 \sqrt{E}}$. 
Recall that $s_{\max}$ linearly increases with $D$, which indicates that
\begin{align}
    \lim_{D \to \infty} \arctan(s_{\max}) = \frac{\pi}{2}.
\end{align}
Therefore, the following limit can be obtained:
\begin{align}
    \lim_{D \to \infty} \frac{2}{D \sqrt{\beta E}} \arctan \left( \frac{\sqrt{\beta} D}{2\sqrt{E}} \right) 
    & = \frac{2}{D \sqrt{\beta E}} \cdot \frac{\pi}{2} \notag\\
    & = \frac{\pi}{D \sqrt{\beta E}}.
\end{align}
Notice that as $D \to \infty$, this term approaches zero.
Therefore, by plugging this into \eqref{eqn: average data rate loss 2}, the following limit can be obtained
\begin{align}
    \lim_{D \to \infty} \mathbb{E}_{\psib}[\Delta R]  = \frac{\alpha^2}{\beta \ln 2} (1 - 0)  = \frac{\alpha^2}{\beta \ln 2}.
\end{align}
The proof of the corollary is complete.
\hfill $\blacksquare$

\end{appendices}




\end{document}